\documentclass[aps,prl,twocolumn,showpacs,superscriptaddress]{revtex4-1}

\usepackage[pdftex]{graphicx} 
\usepackage{natbib}
\usepackage{booktabs}
\usepackage{color}

\newcommand{\one}{Fig.~\ref{f1}}

\newcommand{\three}{Fig.~\ref{f3}}
\newcommand{\four}{Fig.~\ref{f4}}
\newcommand{\five}{Sr$_3$Ir$_2$O$_7$} 
\newcommand{\six}{(Sr$_{1-x}$La$_x$)$_3$Ir$_2$O$_7$}

\begin{document}
\title{A weakly correlated Fermi liquid state with a small Fermi surface in lightly doped {\five}}

\author{A. de la Torre}
\affiliation{Department of Quantum Matter Physics, 24 Quai Ernest-Ansermet, 1211 Geneva 4, Switzerland}
\author{E. C. Hunter}
\affiliation{School of Physics and Astronomy, The University of Edinburgh, James Clerk Maxwell Building, Mayfield Road, Edinburgh EH9 2TT, United Kingdom}
\author{A. Subedi}
\affiliation{Centre de Physique Th\'eorique, \'Ecole Polytechnique, CNRS, 91128 Palaiseau Cedex, France}
\author{S. McKeown Walker}
\affiliation{Department of Quantum Matter Physics, 24 Quai Ernest-Ansermet, 1211 Geneva 4, Switzerland}
\author{A. Tamai}
\affiliation{Department of Quantum Matter Physics, 24 Quai Ernest-Ansermet, 1211 Geneva 4, Switzerland}
\author{T. K. Kim}
\affiliation{Diamond Light Source, Harwell Campus, Didcot, United Kingdom}
\author{M. Hoesch}
\affiliation{Diamond Light Source, Harwell Campus, Didcot, United Kingdom}
\author{ R. S. Perry}
\affiliation{London Centre for Nanotechnology and UCL Centre for Materials Discovery, University College London, London WC1E 6BT, United Kingdom}
\author{A. Georges}
\affiliation{Coll\`ege de France, 11 place Marcelin Berthelot, 75005 Paris, France}
\affiliation{Centre de Physique Th\'eorique, \'Ecole Polytechnique, CNRS, 91128 Palaiseau Cedex, France}
\affiliation{Department of Quantum Matter Physics, 24 Quai Ernest-Ansermet, 1211 Geneva 4, Switzerland}
\author{F. Baumberger}
\affiliation{Department of Quantum Matter Physics, 24 Quai Ernest-Ansermet, 1211 Geneva 4, Switzerland}
\affiliation{Swiss Light Source, Paul Scherrer Institut, CH-5232 Villigen PSI, Switzerland}
\affiliation{SUPA, School of Physics and Astronomy, University of St Andrews, St Andrews, Fife KY16 9SS, United Kingdom}

\date{\today}

\begin{abstract}
We characterize the electron doping evolution of {\six} by means of angle-resolved photoemission. Concomitant with the metal insulator transition around $x\approx0.05$ we find the emergence of coherent quasiparticle states forming a closed small Fermi surface of volume $3x/2$, where $x$ is the independently measured La concentration. The quasiparticle weight $Z$ remains large along the entire Fermi surface, consistent with the moderate renormalization of the low-energy dispersion. This indicates a conventional, weakly correlated Fermi liquid state with a momentum independent residue $Z\approx0.5$ in lightly doped {\five}.
\end{abstract}

\maketitle

The 5$d$ transition metal iridium oxides may host exotic quantum phases emerging from the interplay of correlations and strong spin-orbit coupling~\cite{Shitade2009,Machida2010}. 
Iridates of the Ruddlesden Popper series Sr$_{n+1}$Ir$_n$O$_{3n+1}$ share key structural and electronic properties with the parent compounds of copper oxide superconductors~\cite{Kim2008,Jackeli2009,Kim2012c} and are thus of particular interest as candidate materials for engineering unconventional superconductivity~\cite{Wang2011,Watanabe2013}. 
Despite their partially filled $5d$-shell with a single hole per Ir in the $t_{2g}$ manifold, the layered $n=1,2$ members are antiferromagnetic insulators. This has been attributed to a reduced orbital degeneracy and one-electron bandwidth resulting from strong spin-orbit coupling and structural distortions leading to a single, narrow $j_{\rm{eff}}=1/2$ band that is susceptible to the moderate electron correlations in the Ir $t_{2g}$ shell~\cite{Kim2008,Kim2009, MorettiSala2014}. This picture is supported by band structure calculations for single layer Sr$_2$IrO$_4$. Its low-energy electronic structure can indeed be approximated by a single $j_{\rm{eff}}=1/2$ band which opens a gap in the presence of electron correlations~\cite{Martins2011,Zhang2013}. The resulting insulating state shows in-plane ordered moments and the characteristic spin dynamics of a Heisenberg antiferromagnet with a gapless excitation spectrum~\cite{Jackeli2009,Kim2012c,Fujiyama2012b,Boseggia2013}. The effective low-energy physics can thus be described in a Hubbard model with exchange interaction comparable to cuprates but an electron like non-interacting Fermi surface centered at $\Gamma$~\cite{Jackeli2009,Watanabe2013}. Intriguingly, a recent Monte-Carlo study of this Hamiltonian predicts $d$-wave superconductivity for electron doping but not for hole doping~\cite{Watanabe2013}.
Moreover, recent angle-resolved photoemission (ARPES) data from lightly doped Sr$_2$IrO$_4$~\cite{Kim2014,Cao2014} reproduced much of the unique phenomenology observed in underdoped cuprates, including open Fermi arcs and a temperature dependent pseudogap~\cite{Norman1998,Shen2005,Damascelli2003}. Evidence for strong correlations were also reported for Ru-doped \five~\cite{Dhital2014}.

While the effective single band pseudospin-$1/2$ Mott picture is well established for Sr$_2$IrO$_4$~\cite{Martins2011,Zhang2013}, the properties of the bilayer compound are more ambiguous. On one hand, the magnetic structure of {\five} provides evidence for a $j_{\rm{eff}}=1/2$ state. Both, the out-of-plane collinear ordering and the large spin-wave gap follow naturally from the pseudo-dipolar interaction characteristic of the $j_{\rm{eff}}=1/2$ state~\cite{Boseggia2012,Kim2012,Kim2012b}. 
On the other hand, band structure calculations show mixing of $j_{\rm{eff}}=1/2$ and $3/2$ states in {\five} and a semi metallic electronic structure with a direct band gap throughout the Brillouin zone but band overlap from different $k$-points~\cite{Zhang2013,Okada2013,Moon2008}, qualitatively different from the single particle low-energy electronic structure of Sr$_2$IrO$_4$. 
Experimental evidence for a departure from the ideal $j_{\rm{eff}}=1/2$ Mott state comes from the small ordered moment~\cite{Fujiyama2012}, the small, possibly indirect, charge gap~\cite{Moon2008,Okada2013,Park2014} and the significant overlap in energy between $j_{\rm{eff}}=1/2$ and $3/2$ states found in previous ARPES studies on undoped {\five}~\cite{Wojek2012,King2013,Wang2013,Moreschini2014}. A paramagnetic metallic ground state was reported in \six{} for La concentrations above $x \approx 0.03$~\cite{Li2013}. Yet, little is known about the microscopic electronic structure of doped {\five}.

\begin{figure*}[!ht]
\includegraphics[width=15cm]{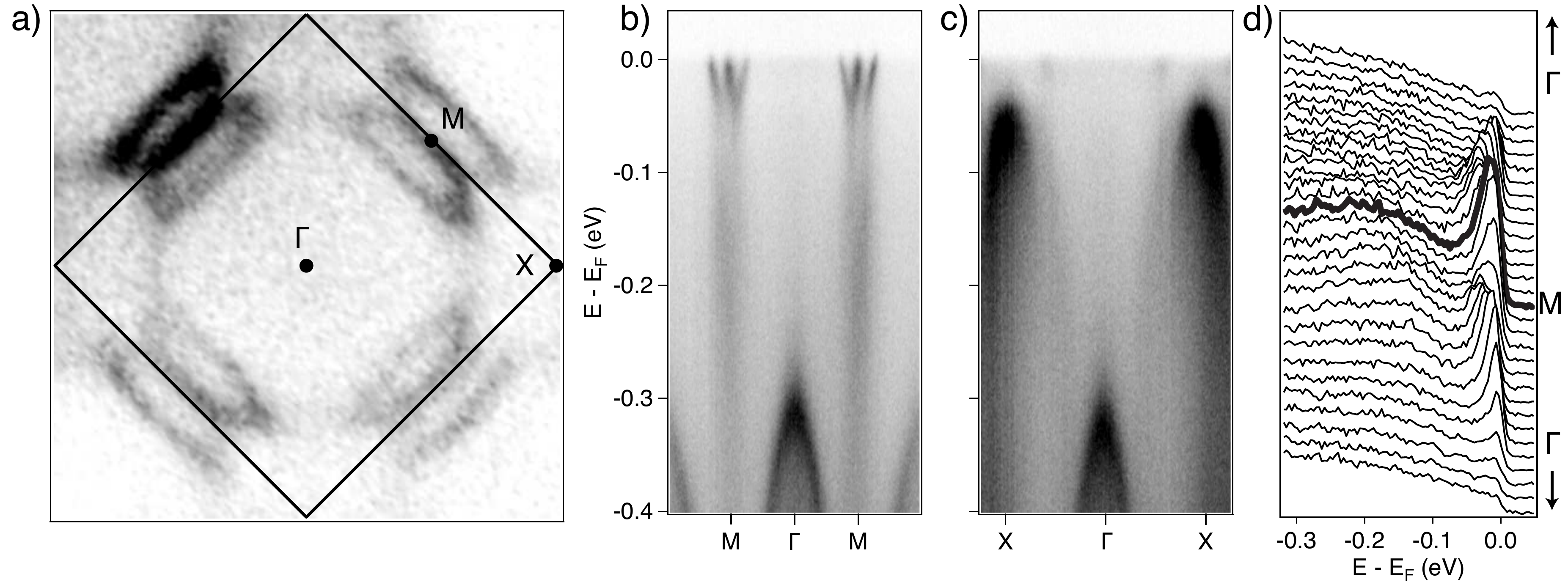} %
\caption{Low energy excitations in lightly doped {\five}. a) Experimental Fermi Surface of (Sr$_{0.935}$La$_{0.065}$)$_3$Ir$_2$O$_7$ taken with $h\nu=83$~eV at 8~K. We denote the $(\pi,0)$ point of the two-dimensional Ir sublattice by X and the $(\pi/2,\pi/2)$ point by M.
b,c) Photoemission intensity along the $\Gamma$M and $\Gamma$X high symmetry directions, respectively. d) Stack of energy distribution curves (EDCs) around the Fermi surface extracted from the data shown in panel b).}
\label{f1}
\end{figure*}

Here we show that lightly doped {\five} supports a conventional, weakly-correlated Fermi-liquid-like metallic state, strikingly different from the electronic structure of underdoped cuprates and from recent ARPES results from doped Sr$_2$IrO$_4$~\cite{Kim2014,Cao2014}. 
\\

ARPES measurements were performed at the I05 beamline of Diamond Light Source. Crystals of {\six} were flux grown and showed a paramagnetic metallic ground state for $x>0.045$. Details of the sample growth and characterization will be presented elsewhere~\cite{Hunter2014}.
The samples were cleaved at pressures $<10^{-10}$~mbar and temperatures $<50$~K. Measurements were made using photon energies between 30~eV and 120~eV. All presented data was acquired at 83~eV with an energy resolution of 15~meV. The sample temperature was 8~K and 50~K for conducting and insulating samples, respectively. 
The electronic structure calculations were performed using the generalized full-potential method within local density approximation (LDA) as implemented in the WIEN2k software package~\cite{Blaha2001}. Doping was treated within the virtual crystal approximation (VCA).
A value of $U$ = 3 eV was used for the on-site Coulomb repulsion. We used the $Bbcb$ space group with lattice parameters $a = 5.517, b = 5.515, c = 20.897$~\AA~\cite{Hunter2014} and relaxed the internal atomic positions for each value of doping.


{\one} shows the the main features of the low energy electronic structure of electron doped {\six} for $x=0.065$ where the samples are metallic down to low temperatures~\cite{Hunter2014,Li2013}. The spectral weight at the Fermi level is concentrated along the Brillouin zone diagonal and strongly suppressed along the Ir -- Ir nearest neighbor direction, reminiscent of the nodal / antinodal dichotomy in lightly doped cuprates~\cite{Damascelli2003,Shen2005} and the recent observation of Fermi arcs in doped Sr$_2$IrO$_4$~\cite{Kim2014,Cao2014}. Despite not having any special meaning for the iridates, we use the term "nodal" throughout this manuscript to facilitate the comparison with the copper oxides. The dispersion plots and energy distribution curves (EDCs) shown in Fig.~1b-d) confirm this picture. We find coherent quasiparticle-like states along the nodal cut but only faint low-energy spectral weight with a broader momentum distribution along the antipodal direction $\Gamma$X. The intense hole-like bands observed at higher energy ($-0.3$~eV at $\Gamma$ and $-0.05$~eV at X) originate from the occupied $j_{\rm{eff}}=3/2$ and $1/2$ states in the undoped parent compound, respectively~\cite{King2013,Zhang2013}. Along the nodal direction, where the quasiparticle states are clearly separated from the bands of the parent compound, we observe incoherent tails with nearly vertical dispersion separated from the quasiparticle bands by a slight suppression of spectral weight resulting in a peak-dip-hump line shape indicative of coupling to a bosonic mode.
While the distribution of spectral weight alone does not permit a reliable determination of the coupling strength, the dominant mode energy can be estimated from the separation of peak and hump to be around 100 - 150~meV and thus in the range where the magnon density of states peaks~\cite{Kim2012b}.

\begin{figure*}[!ht]
\includegraphics[width=17.5cm]{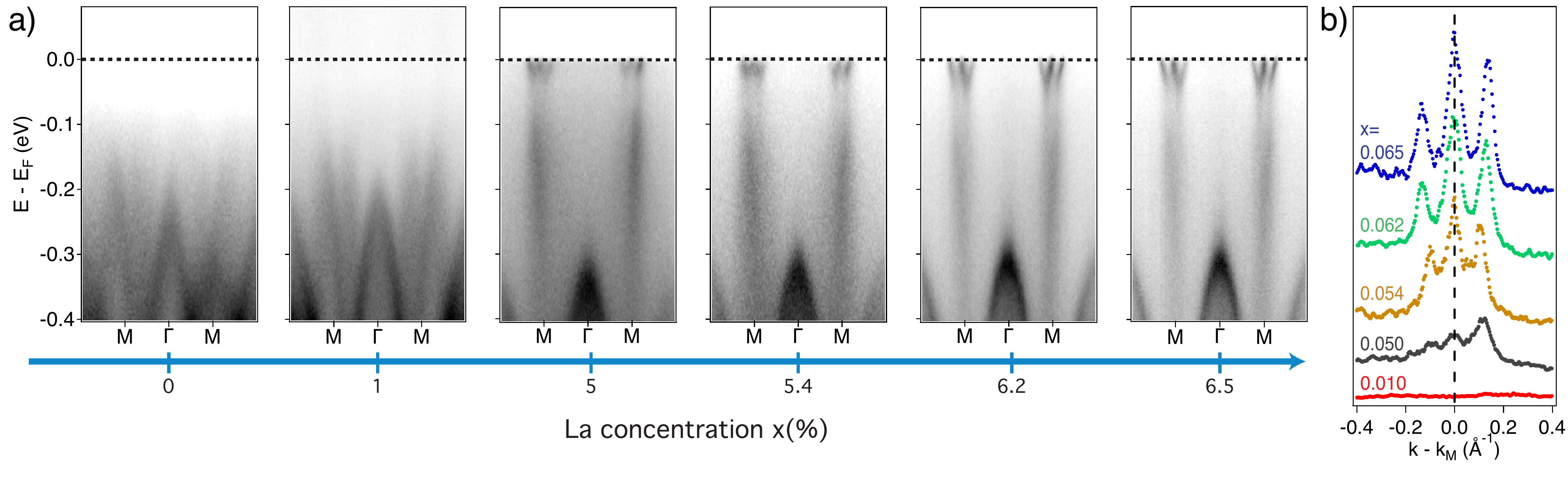} %
\caption{Doping evolution of {\six}. a) Dispersion along the nodal $\Gamma$M direction. Coherent quasiparticle states emerge for $x>0.05$. b) MDCs at the Fermi energy. The Fermi wave vector increases monotonically with doping, indicative of an increasing density of itinerant carriers.}
\label{f2}
\end{figure*}

The systematic evolution of the electronic structure of {\six} with increasing doping is shown in Fig.~2. For $x=0.01$, faint spectral weight appears around the M-point in the gap of the parent compound, indicating localized carriers, consistent with the insulating ground state of these samples. At $x=0.05$, coincident with the metal insulator transition in transport, we observe the emergence of coherent quasiparticle states whose Fermi wave vector and band width increase monotonically with doping. We note that the doping level at which the first quasiparticle-like states appear in {\five}, and their nodal Fermi velocity of $\approx10^5$~m/s are comparable to lightly doped cuprates~\cite{Ino2002,Shen2004,Shen2005,Yoshida2006}. However, the quasiparticle pole is more pronounced here and more clearly separated from the incoherent weight at higher energy.

\begin{figure}[htb]
\centering
\includegraphics[width=7.5cm]{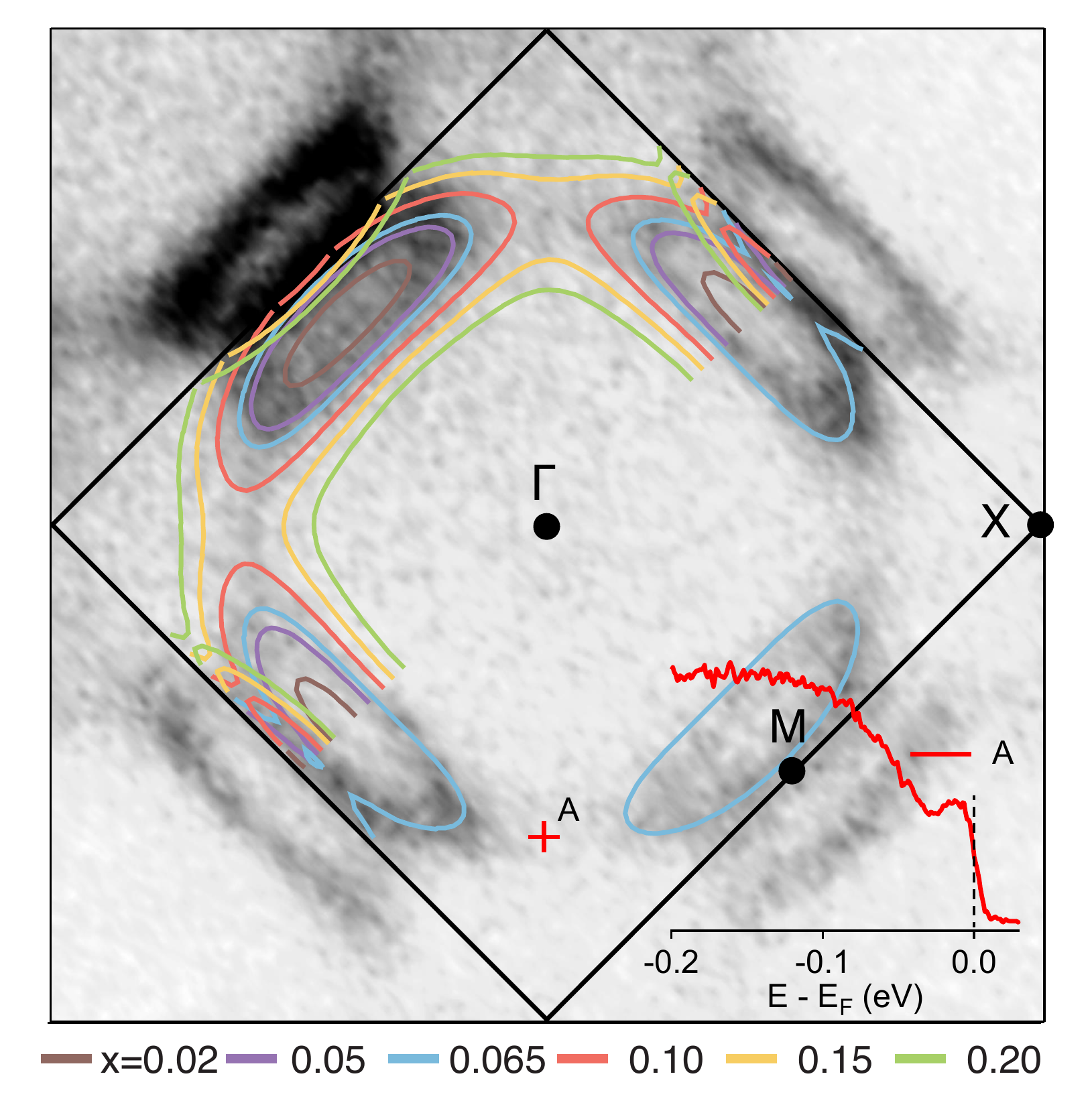} %
\caption{LDA+U calculations of the Fermi surface for $U=3$~eV and different doping levels treated by the VCA superimposed on ARPES data for $x=0.065$. The inset shows an EDC taken at 8~K at $\approx(0,-0.6\pi)$ demonstrating the absence of a pseudogap in the antinodal direction at the measurement temperature of 8~K.}
\label{f3}
\end{figure}

\begin{figure}[htb]
\centering
\includegraphics[width=8.44cm]{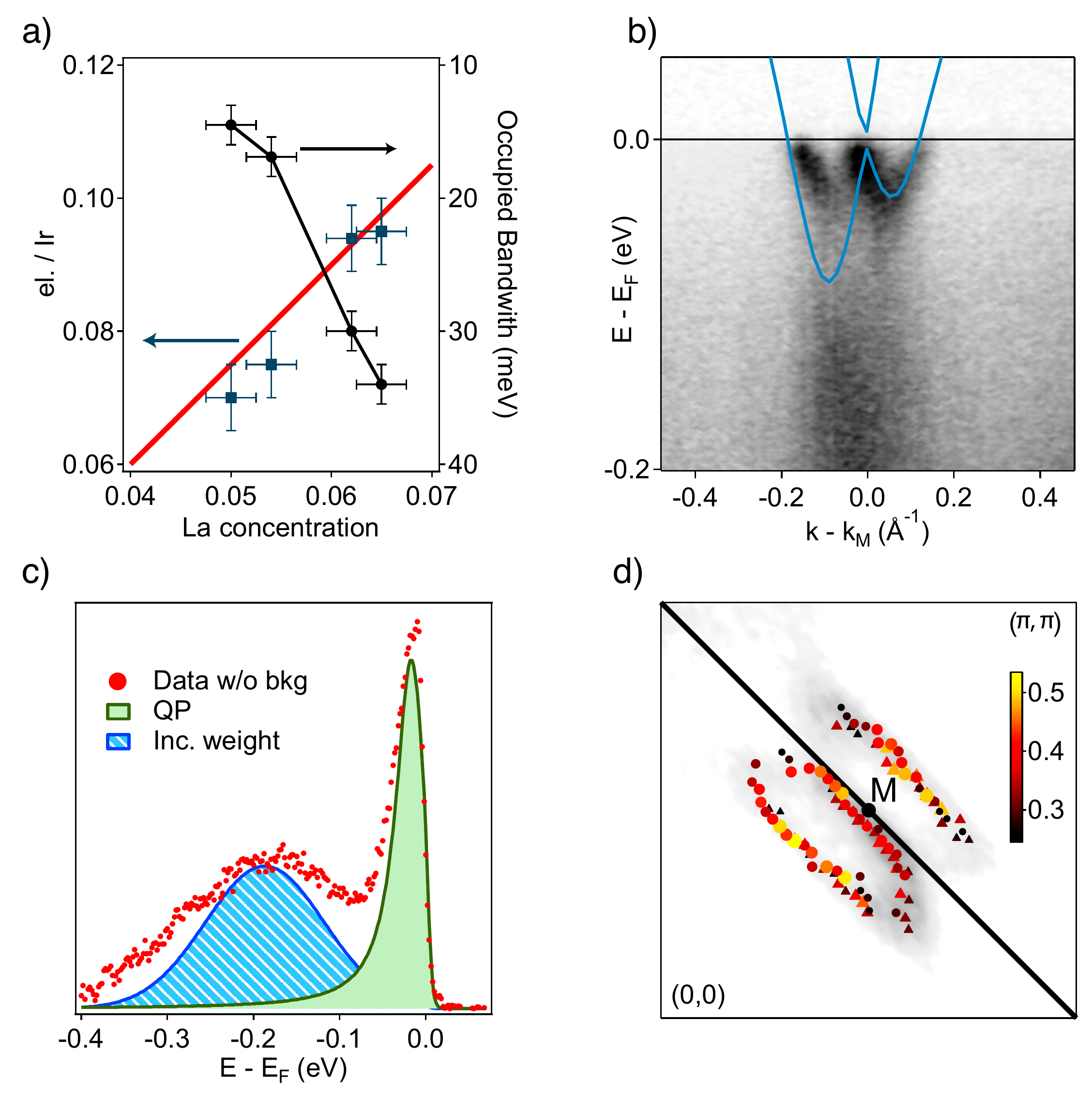} %
\caption{Luttinger volume and quasiparticle residue.
a) Luttinger volume of the experimental Fermi surface (blue), compared with the nominal Luttinger volume of $3/2x$ arising from the stoichiometry of {\six} if every Lanthanum atom dopes one itinerant electron (red line). The doping evolution of the conduction band minimum is shown in black on the right axis.
b) Band dispersion along $\Gamma$M together with a LDA-VCA calculation for $x=0.065$ (U=0). The hole pocket found in this calculation has been suppressed for clarity and the electron pocket at M was shifted in energy to match the experimental Fermi wave vectors.
c,d) Quasiparticle residue $Z$ along the Fermi surfaces at $\pm(\pi/2,\pi/2)$ (circles and triangles, respectively) estimated from the weight of the coherent quasiparticles.}
\label{f4}
\end{figure}

For the remainder of this paper, we discuss the origin of the phenomenology described above.
To this end, we first compare our data with band structure calculations. Using the local density approximation (LDA), we find a metallic ground state of undoped \five, consistent with earlier reports~\cite{Moon2008,Okada2013}. The doubling of the in-plane unit cell due to the rotations of the IrO$_6$ octahedra around the c-axis and the bilayer nature of \five{} causes an even number of Ir 5d$^5$ ions per unit cell and thus an even number of electrons. Yet, the significant itineracy of the states prevents the formation of a band insulator and leads to a compensated metallic state with a small hole pocket at $\Gamma$ and electron pockets at the M-points. 
The former is rapidly pushed below the Fermi level by adding correlations within LDA+U, eventually reproducing the insulating ground state of {\five}~\cite{Moon2008,Okada2013}. In {\three} we superimpose LDA+U Fermi surfaces for different doping levels treated within the virtual crystal approximation (VCA) on data taken for $x=0.065$.
The good agreement of these calculations with the data strongly suggests that the nodal/antinodal dichotomy is a band structure effect and thus of fundamentally different origin than in the cuprates. This interpretation is supported by the line shape of spectra around $(0,-0.6\pi)$, where the intensity at the Fermi level peaks along $\Gamma$X (see insert to {\three}). The small but distinct Fermi step is inconsistent with the formation of a pseudogap driving the dichotomy of the spectral weight as reported for doped Sr$_2$IrO$_4$~\cite{Kim2014}. Instead, the reduced weight along $\Gamma$X can be understood as the tail of a pole in the spectral function above the chemical potential arising from the merging of electron pockets from different M-points, as seen in the LDA+U calculations for $x\approx0.15$. Hence, we conclude that La doped {\five} shows closed electron-like Fermi pockets with a double-lens shape.

The Fermi surface pockets are significantly smaller than half of the Brillouin zone, rendering the Fermiology of doped {\five} qualitatively different from cuprates, which show open Fermi arcs at low doping, evolving into a {\it large} Fermi surface of volume $(1+x)$ near optimal doping~\cite{Damascelli2003,Shen2005}. We have extracted the Luttinger volume in {\six} from extensive fits to the experimental Fermi surface. As shown in {\four}a), the values are fully consistent with a {\it small} Fermi surface of volume $3/2\:x$, where $x$ is the independently measured La concentration. The factor of 3/2 arises from the stoichiometry of {\six} and indicates that every La ion contributes one electron to the Fermi sea.
The interpretation of our data within a conventional Fermi liquid picture with small Fermi surfaces is supported by a calculation of the electronic specific heat from the ARPES data. Using the experimental Fermi surface volume $A_{\rm{FS}}$ and quasiparticle velocities for $x=0.065$ we calculate cyclotron masses $m^*=\hbar^2\: \partial A_{\rm{FS}}/2\pi \partial\epsilon= 3.1(9)$~m$_e$ for each Fermi surface lens. The doping dependence of $m^*$ is negligible within the current accuracy of our experiment. Assuming two-dimensionality, these masses correspond to a Sommerfeld coefficient $\gamma^{\rm{ARPES}}=9(2)$~mJ/molK$^2$, in good agreement with the direct thermodynamic measurement of $\gamma=8.4-11.5$~mJ/molK$^2$ for different metallic samples, respectively~\cite{Hunter2014}.
This demonstrates that all low-energy charge excitations arise from the small Fermi surface pockets identified above. Furthermore, it strongly suggests that our ARPES data represents the bulk electronic structure.

The role of correlations in the low-energy excitations can be estimated from the quasiparticle weight and the renormalization of the low-energy dispersion. For a momentum independent self-energy, the quasiparticle residue $Z=v_{\rm{F}}/v_{\rm{bare}}$. While $v_{\rm{bare}}$ cannot be determined directly from experiment, it can often be approximated by band structure calculations within the LDA. In {\four}b we thus compare the experimental quasiparticle dispersion for $x=0.065$ with an LDA-VCA calculation for the same doping and a fully relaxed crystal structure. 
Using averaged Fermi velocities from all Fermi surface crossings we find $Z=0.5(2)$. The error bar is estimated from a slight uncertainty in the experimental Fermi velocity due to finite resolution effects and the high sensitivity of the calculated dispersion to the rotation angle of the octahedra.
As an alternative measure of the quasiparticle residue $Z$, we analyze the coherent weight. To this end we subtract a smooth background from the raw EDCs and fit the resulting spectra along the entire Fermi surface with two peaks representing the coherent quasiparticle and incoherent hump (see {\four}c)). While this analysis is somewhat model dependent and tends to underestimate $Z$, it clearly shows a significant coherent quasiparticle weight $Z=0.25$ -- 0.5 along the entire Fermi surface, in fair agreement with the moderate renormalization of the dispersion but fundamentally different from the case of lightly doped cuprates~\cite{Shen2004}.

The role of correlations in {\five} is thus rather intricate. On one hand, they are crucial for driving the insulating ground state in the parent compound via a substantial orbital dependent shift and deformation of the bare band structure~\cite{Martins2011,Zhang2013}. On the other hand, they appear to play a minor role in the low-energy physics once metallicity is induced by light electron doping with La.
Both the large weight of the coherent quasiparticle peaks and the low mass enhancement in electron doped {\five} are not only clearly different from cuprates but also from the isostructural strongly correlated ruthenate Sr$_3$Ru$_2$O$_7$~\cite{Tamai2008} and more akin to the robust 4d$^5$ metal Sr$_2$RhO$_4$~\cite{Baumberger2006}. This confirms the important role of the average occupation of the $d$-shell observed in recent DMFT studies~\cite{Georges2013}.

In conclusion, we observed a phenomenology of the spectral function in {\six} reminiscent of lightly doped cuprates. However, we argue that this behavior can be understood in a simple band structure picture. In particular, we show that our data is consistent with a small closed Fermi surface with a volume corresponding to the doping $x$, in stark contrast to lightly doped cuprates~\cite{Norman1998,Shen2005,Damascelli2003} and surface doped Sr$_2$IrO$_4$~\cite{Kim2014}. The light quasiparticle states with large coherent weight observed in metallic {\six} are characteristic of a weakly correlated Fermi liquid. Together, these findings suggest that the properties of {\five} are unique among doped correlated insulators and not suitable for the engineering of cuprate-like high-T$_{\rm{c}}$ superconductivity. It remains an open question how our findings relate to recent reports of strongly correlated states in doped single layer iridates~\cite{Kim2014,Cao2014} and Ru substituted bilayer iridates~\cite{Dhital2014}.

\begin{acknowledgments}
We gratefully acknowledge discussions with D.F. McMorrow. This work was funded by the Swiss National Science Foundation (200021-146995), the UK Royal Society, the UK-EPSRC and the European Research Council (ERC-319286 QMAC).
\end{acknowledgments}

\bibliography{Bib_v3.bib}

\begin{thebibliography}{39}%
\makeatletter
\providecommand \@ifxundefined [1]{%
 \@ifx{#1\undefined}
}%
\providecommand \@ifnum [1]{%
 \ifnum #1\expandafter \@firstoftwo
 \else \expandafter \@secondoftwo
 \fi
}%
\providecommand \@ifx [1]{%
 \ifx #1\expandafter \@firstoftwo
 \else \expandafter \@secondoftwo
 \fi
}%
\providecommand \natexlab [1]{#1}%
\providecommand \enquote  [1]{``#1''}%
\providecommand \bibnamefont  [1]{#1}%
\providecommand \bibfnamefont [1]{#1}%
\providecommand \citenamefont [1]{#1}%
\providecommand \href@noop [0]{\@secondoftwo}%
\providecommand \href [0]{\begingroup \@sanitize@url \@href}%
\providecommand \@href[1]{\@@startlink{#1}\@@href}%
\providecommand \@@href[1]{\endgroup#1\@@endlink}%
\providecommand \@sanitize@url [0]{\catcode `\\12\catcode `\$12\catcode
  `\&12\catcode `\#12\catcode `\^12\catcode `\_12\catcode `\%12\relax}%
\providecommand \@@startlink[1]{}%
\providecommand \@@endlink[0]{}%
\providecommand \url  [0]{\begingroup\@sanitize@url \@url }%
\providecommand \@url [1]{\endgroup\@href {#1}{\urlprefix }}%
\providecommand \urlprefix  [0]{URL }%
\providecommand \Eprint [0]{\href }%
\providecommand \doibase [0]{http://dx.doi.org/}%
\providecommand \selectlanguage [0]{\@gobble}%
\providecommand \bibinfo  [0]{\@secondoftwo}%
\providecommand \bibfield  [0]{\@secondoftwo}%
\providecommand \translation [1]{[#1]}%
\providecommand \BibitemOpen [0]{}%
\providecommand \bibitemStop [0]{}%
\providecommand \bibitemNoStop [0]{.\EOS\space}%
\providecommand \EOS [0]{\spacefactor3000\relax}%
\providecommand \BibitemShut  [1]{\csname bibitem#1\endcsname}%
\let\auto@bib@innerbib\@empty
\bibitem [{\citenamefont {Shitade}\ \emph {et~al.}(2009)\citenamefont
  {Shitade}, \citenamefont {Katsura}, \citenamefont {Kune\v{s}}, \citenamefont
  {Qi}, \citenamefont {Zhang},\ and\ \citenamefont {Nagaosa}}]{Shitade2009}%
  \BibitemOpen
  \bibfield  {author} {\bibinfo {author} {\bibfnamefont {A.}~\bibnamefont
  {Shitade}}, \bibinfo {author} {\bibfnamefont {H.}~\bibnamefont {Katsura}},
  \bibinfo {author} {\bibfnamefont {J.}~\bibnamefont {Kune\v{s}}}, \bibinfo
  {author} {\bibfnamefont {X.-L.}\ \bibnamefont {Qi}}, \bibinfo {author}
  {\bibfnamefont {S.-C.}\ \bibnamefont {Zhang}}, \ and\ \bibinfo {author}
  {\bibfnamefont {N.}~\bibnamefont {Nagaosa}},\ }\href {\doibase
  10.1103/PhysRevLett.102.256403} {\bibfield  {journal} {\bibinfo  {journal}
  {Phys. Rev. Lett.}\ }\textbf {\bibinfo {volume} {102}},\ \bibinfo {pages}
  {256403} (\bibinfo {year} {2009})}\BibitemShut {NoStop}%
\bibitem [{\citenamefont {Machida}\ \emph {et~al.}(2010)\citenamefont
  {Machida}, \citenamefont {Nakatsuji}, \citenamefont {Onoda}, \citenamefont
  {Tayama},\ and\ \citenamefont {Sakakibara}}]{Machida2010}%
  \BibitemOpen
  \bibfield  {author} {\bibinfo {author} {\bibfnamefont {Y.}~\bibnamefont
  {Machida}}, \bibinfo {author} {\bibfnamefont {S.}~\bibnamefont {Nakatsuji}},
  \bibinfo {author} {\bibfnamefont {S.}~\bibnamefont {Onoda}}, \bibinfo
  {author} {\bibfnamefont {T.}~\bibnamefont {Tayama}}, \ and\ \bibinfo {author}
  {\bibfnamefont {T.}~\bibnamefont {Sakakibara}},\ }\href {\doibase
  10.1038/nature08680} {\bibfield  {journal} {\bibinfo  {journal} {Nature}\
  }\textbf {\bibinfo {volume} {463}},\ \bibinfo {pages} {210} (\bibinfo {year}
  {2010})}\BibitemShut {NoStop}%
\bibitem [{\citenamefont {Kim}\ \emph {et~al.}(2008)\citenamefont {Kim},
  \citenamefont {Jin}, \citenamefont {Moon}, \citenamefont {Kim}, \citenamefont
  {Park}, \citenamefont {Leem}, \citenamefont {Yu}, \citenamefont {Noh},
  \citenamefont {Kim}, \citenamefont {Oh}, \citenamefont {Park}, \citenamefont
  {Durairaj}, \citenamefont {Cao},\ and\ \citenamefont {Rotenberg}}]{Kim2008}%
  \BibitemOpen
  \bibfield  {author} {\bibinfo {author} {\bibfnamefont {B.~J.}\ \bibnamefont
  {Kim}}, \bibinfo {author} {\bibfnamefont {H.}~\bibnamefont {Jin}}, \bibinfo
  {author} {\bibfnamefont {S.~J.}\ \bibnamefont {Moon}}, \bibinfo {author}
  {\bibfnamefont {J.~Y.}\ \bibnamefont {Kim}}, \bibinfo {author} {\bibfnamefont
  {B.~G.}\ \bibnamefont {Park}}, \bibinfo {author} {\bibfnamefont {C.~S.}\
  \bibnamefont {Leem}}, \bibinfo {author} {\bibfnamefont {J.}~\bibnamefont
  {Yu}}, \bibinfo {author} {\bibfnamefont {T.~W.}\ \bibnamefont {Noh}},
  \bibinfo {author} {\bibfnamefont {C.}~\bibnamefont {Kim}}, \bibinfo {author}
  {\bibfnamefont {S.~J.}\ \bibnamefont {Oh}}, \bibinfo {author} {\bibfnamefont
  {J.~H.}\ \bibnamefont {Park}}, \bibinfo {author} {\bibfnamefont
  {V.}~\bibnamefont {Durairaj}}, \bibinfo {author} {\bibfnamefont
  {G.}~\bibnamefont {Cao}}, \ and\ \bibinfo {author} {\bibfnamefont
  {E.}~\bibnamefont {Rotenberg}},\ }\href
  {http://link.aps.org/abstract/PRL/v101/e076402} {\bibfield  {journal}
  {\bibinfo  {journal} {Phys. Rev. Lett.}\ }\textbf {\bibinfo {volume} {101}},\
  \bibinfo {pages} {076402} (\bibinfo {year} {2008})}\BibitemShut {NoStop}%
\bibitem [{\citenamefont {Jackeli}\ and\ \citenamefont
  {Khaliullin}(2009)}]{Jackeli2009}%
  \BibitemOpen
  \bibfield  {author} {\bibinfo {author} {\bibfnamefont {G.}~\bibnamefont
  {Jackeli}}\ and\ \bibinfo {author} {\bibfnamefont {G.}~\bibnamefont
  {Khaliullin}},\ }\href {\doibase 10.1103/PhysRevLett.102.017205} {\bibfield
  {journal} {\bibinfo  {journal} {Phys. Rev. Lett.}\ }\textbf {\bibinfo
  {volume} {102}},\ \bibinfo {pages} {017205} (\bibinfo {year}
  {2009})}\BibitemShut {NoStop}%
\bibitem [{\citenamefont {Kim}\ \emph {et~al.}(2012{\natexlab{a}})\citenamefont
  {Kim}, \citenamefont {Casa}, \citenamefont {Upton}, \citenamefont {Gog},
  \citenamefont {Kim}, \citenamefont {Mitchell}, \citenamefont {van
  Veenendaal}, \citenamefont {Daghofer}, \citenamefont {van~den Brink},
  \citenamefont {Khaliullin},\ and\ \citenamefont {Kim}}]{Kim2012c}%
  \BibitemOpen
  \bibfield  {author} {\bibinfo {author} {\bibfnamefont {J.}~\bibnamefont
  {Kim}}, \bibinfo {author} {\bibfnamefont {D.}~\bibnamefont {Casa}}, \bibinfo
  {author} {\bibfnamefont {M.~H.}\ \bibnamefont {Upton}}, \bibinfo {author}
  {\bibfnamefont {T.}~\bibnamefont {Gog}}, \bibinfo {author} {\bibfnamefont
  {Y.-J.}\ \bibnamefont {Kim}}, \bibinfo {author} {\bibfnamefont {J.~F.}\
  \bibnamefont {Mitchell}}, \bibinfo {author} {\bibfnamefont {M.}~\bibnamefont
  {van Veenendaal}}, \bibinfo {author} {\bibfnamefont {M.}~\bibnamefont
  {Daghofer}}, \bibinfo {author} {\bibfnamefont {J.}~\bibnamefont {van~den
  Brink}}, \bibinfo {author} {\bibfnamefont {G.}~\bibnamefont {Khaliullin}}, \
  and\ \bibinfo {author} {\bibfnamefont {B.~J.}\ \bibnamefont {Kim}},\ }\href
  {\doibase 10.1103/PhysRevLett.108.177003} {\bibfield  {journal} {\bibinfo
  {journal} {Phys. Rev. Lett.}\ }\textbf {\bibinfo {volume} {108}},\ \bibinfo
  {pages} {177003} (\bibinfo {year} {2012}{\natexlab{a}})}\BibitemShut
  {NoStop}%
\bibitem [{\citenamefont {Wang}\ and\ \citenamefont
  {Senthil}(2011)}]{Wang2011}%
  \BibitemOpen
  \bibfield  {author} {\bibinfo {author} {\bibfnamefont {F.}~\bibnamefont
  {Wang}}\ and\ \bibinfo {author} {\bibfnamefont {T.}~\bibnamefont {Senthil}},\
  }\href {http://link.aps.org/doi/10.1103/PhysRevLett.106.136402} {\bibfield
  {journal} {\bibinfo  {journal} {Phys. Rev. Lett.}\ }\textbf {\bibinfo
  {volume} {106}},\ \bibinfo {pages} {136402} (\bibinfo {year}
  {2011})}\BibitemShut {NoStop}%
\bibitem [{\citenamefont {Watanabe}\ \emph {et~al.}(2013)\citenamefont
  {Watanabe}, \citenamefont {Shirakawa},\ and\ \citenamefont
  {Yunoki}}]{Watanabe2013}%
  \BibitemOpen
  \bibfield  {author} {\bibinfo {author} {\bibfnamefont {H.}~\bibnamefont
  {Watanabe}}, \bibinfo {author} {\bibfnamefont {T.}~\bibnamefont {Shirakawa}},
  \ and\ \bibinfo {author} {\bibfnamefont {S.}~\bibnamefont {Yunoki}},\ }\href
  {\doibase 10.1103/PhysRevLett.110.027002} {\bibfield  {journal} {\bibinfo
  {journal} {Phys. Rev. Lett.}\ }\textbf {\bibinfo {volume} {110}},\ \bibinfo
  {pages} {027002} (\bibinfo {year} {2013})}\BibitemShut {NoStop}%
\bibitem [{\citenamefont {Kim}\ \emph {et~al.}(2009)\citenamefont {Kim},
  \citenamefont {Ohsumi}, \citenamefont {Komesu}, \citenamefont {Sakai},
  \citenamefont {Morita}, \citenamefont {Takagi},\ and\ \citenamefont
  {Arima}}]{Kim2009}%
  \BibitemOpen
  \bibfield  {author} {\bibinfo {author} {\bibfnamefont {B.~J.}\ \bibnamefont
  {Kim}}, \bibinfo {author} {\bibfnamefont {H.}~\bibnamefont {Ohsumi}},
  \bibinfo {author} {\bibfnamefont {T.}~\bibnamefont {Komesu}}, \bibinfo
  {author} {\bibfnamefont {S.}~\bibnamefont {Sakai}}, \bibinfo {author}
  {\bibfnamefont {T.}~\bibnamefont {Morita}}, \bibinfo {author} {\bibfnamefont
  {H.}~\bibnamefont {Takagi}}, \ and\ \bibinfo {author} {\bibfnamefont
  {T.}~\bibnamefont {Arima}},\ }\href
  {http://www.sciencemag.org/cgi/content/abstract/323/5919/1329} {\bibfield
  {journal} {\bibinfo  {journal} {Science}\ }\textbf {\bibinfo {volume}
  {323}},\ \bibinfo {pages} {1329} (\bibinfo {year} {2009})}\BibitemShut
  {NoStop}%
\bibitem [{\citenamefont {{Moretti Sala}}\ \emph {et~al.}(2014)\citenamefont
  {{Moretti Sala}}, \citenamefont {Boseggia}, \citenamefont {McMorrow},\ and\
  \citenamefont {Monaco}}]{MorettiSala2014}%
  \BibitemOpen
  \bibfield  {author} {\bibinfo {author} {\bibfnamefont {M.}~\bibnamefont
  {{Moretti Sala}}}, \bibinfo {author} {\bibfnamefont {S.}~\bibnamefont
  {Boseggia}}, \bibinfo {author} {\bibfnamefont {D.~F.}\ \bibnamefont
  {McMorrow}}, \ and\ \bibinfo {author} {\bibfnamefont {G.}~\bibnamefont
  {Monaco}},\ }\href {\doibase 10.1103/PhysRevLett.112.026403} {\bibfield
  {journal} {\bibinfo  {journal} {Physical Review Letters}\ }\textbf {\bibinfo
  {volume} {112}},\ \bibinfo {pages} {026403} (\bibinfo {year}
  {2014})}\BibitemShut {NoStop}%
\bibitem [{\citenamefont {Martins}\ \emph {et~al.}(2011)\citenamefont
  {Martins}, \citenamefont {Aichhorn}, \citenamefont {Vaugier},\ and\
  \citenamefont {Biermann}}]{Martins2011}%
  \BibitemOpen
  \bibfield  {author} {\bibinfo {author} {\bibfnamefont {C.}~\bibnamefont
  {Martins}}, \bibinfo {author} {\bibfnamefont {M.}~\bibnamefont {Aichhorn}},
  \bibinfo {author} {\bibfnamefont {L.}~\bibnamefont {Vaugier}}, \ and\
  \bibinfo {author} {\bibfnamefont {S.}~\bibnamefont {Biermann}},\ }\href
  {http://link.aps.org/doi/10.1103/PhysRevLett.107.266404} {\bibfield
  {journal} {\bibinfo  {journal} {Phys. Rev. Lett.}\ }\textbf {\bibinfo
  {volume} {107}},\ \bibinfo {pages} {266404} (\bibinfo {year}
  {2011})}\BibitemShut {NoStop}%
\bibitem [{\citenamefont {Zhang}\ \emph {et~al.}(2013)\citenamefont {Zhang},
  \citenamefont {Haule},\ and\ \citenamefont {Vanderbilt}}]{Zhang2013}%
  \BibitemOpen
  \bibfield  {author} {\bibinfo {author} {\bibfnamefont {H.}~\bibnamefont
  {Zhang}}, \bibinfo {author} {\bibfnamefont {K.}~\bibnamefont {Haule}}, \ and\
  \bibinfo {author} {\bibfnamefont {D.}~\bibnamefont {Vanderbilt}},\ }\href
  {\doibase 10.1103/PhysRevLett.111.246402} {\bibfield  {journal} {\bibinfo
  {journal} {Phys. Rev. Lett.}\ }\textbf {\bibinfo {volume} {111}},\ \bibinfo
  {pages} {246402} (\bibinfo {year} {2013})}\BibitemShut {NoStop}%
\bibitem [{\citenamefont {Fujiyama}\ \emph
  {et~al.}(2012{\natexlab{a}})\citenamefont {Fujiyama}, \citenamefont {Ohsumi},
  \citenamefont {Komesu}, \citenamefont {Matsuno}, \citenamefont {Kim},
  \citenamefont {Takata}, \citenamefont {Arima},\ and\ \citenamefont
  {Takagi}}]{Fujiyama2012b}%
  \BibitemOpen
  \bibfield  {author} {\bibinfo {author} {\bibfnamefont {S.}~\bibnamefont
  {Fujiyama}}, \bibinfo {author} {\bibfnamefont {H.}~\bibnamefont {Ohsumi}},
  \bibinfo {author} {\bibfnamefont {T.}~\bibnamefont {Komesu}}, \bibinfo
  {author} {\bibfnamefont {J.}~\bibnamefont {Matsuno}}, \bibinfo {author}
  {\bibfnamefont {B.~J.}\ \bibnamefont {Kim}}, \bibinfo {author} {\bibfnamefont
  {M.}~\bibnamefont {Takata}}, \bibinfo {author} {\bibfnamefont
  {T.}~\bibnamefont {Arima}}, \ and\ \bibinfo {author} {\bibfnamefont
  {H.}~\bibnamefont {Takagi}},\ }\href {\doibase
  10.1103/PhysRevLett.108.247212} {\bibfield  {journal} {\bibinfo  {journal}
  {Phys. Rev. Lett.}\ }\textbf {\bibinfo {volume} {108}},\ \bibinfo {pages}
  {247212} (\bibinfo {year} {2012}{\natexlab{a}})}\BibitemShut {NoStop}%
\bibitem [{\citenamefont {Boseggia}\ \emph {et~al.}(2013)\citenamefont
  {Boseggia}, \citenamefont {Springell}, \citenamefont {Walker}, \citenamefont
  {R{\o}nnow}, \citenamefont {R\"{u}egg}, \citenamefont {Okabe}, \citenamefont
  {Isobe}, \citenamefont {Perry}, \citenamefont {Collins},\ and\ \citenamefont
  {McMorrow}}]{Boseggia2013}%
  \BibitemOpen
  \bibfield  {author} {\bibinfo {author} {\bibfnamefont {S.}~\bibnamefont
  {Boseggia}}, \bibinfo {author} {\bibfnamefont {R.}~\bibnamefont {Springell}},
  \bibinfo {author} {\bibfnamefont {H.~C.}\ \bibnamefont {Walker}}, \bibinfo
  {author} {\bibfnamefont {H.~M.}\ \bibnamefont {R{\o}nnow}}, \bibinfo {author}
  {\bibfnamefont {C.}~\bibnamefont {R\"{u}egg}}, \bibinfo {author}
  {\bibfnamefont {H.}~\bibnamefont {Okabe}}, \bibinfo {author} {\bibfnamefont
  {M.}~\bibnamefont {Isobe}}, \bibinfo {author} {\bibfnamefont {R.~S.}\
  \bibnamefont {Perry}}, \bibinfo {author} {\bibfnamefont {S.~P.}\ \bibnamefont
  {Collins}}, \ and\ \bibinfo {author} {\bibfnamefont {D.~F.}\ \bibnamefont
  {McMorrow}},\ }\href {\doibase 10.1103/PhysRevLett.110.117207} {\bibfield
  {journal} {\bibinfo  {journal} {Phys. Rev. Lett.}\ }\textbf {\bibinfo
  {volume} {110}},\ \bibinfo {pages} {117207} (\bibinfo {year}
  {2013})}\BibitemShut {NoStop}%
\bibitem [{\citenamefont {Kim}\ \emph {et~al.}(2014)\citenamefont {Kim},
  \citenamefont {Krupin}, \citenamefont {Denlinger}, \citenamefont {Bostwick},
  \citenamefont {Rotenberg}, \citenamefont {Zhao}, \citenamefont {Mitchell},
  \citenamefont {Allen},\ and\ \citenamefont {Kim}}]{Kim2014}%
  \BibitemOpen
  \bibfield  {author} {\bibinfo {author} {\bibfnamefont {Y.~K.}\ \bibnamefont
  {Kim}}, \bibinfo {author} {\bibfnamefont {O.}~\bibnamefont {Krupin}},
  \bibinfo {author} {\bibfnamefont {J.~D.}\ \bibnamefont {Denlinger}}, \bibinfo
  {author} {\bibfnamefont {A.}~\bibnamefont {Bostwick}}, \bibinfo {author}
  {\bibfnamefont {E.}~\bibnamefont {Rotenberg}}, \bibinfo {author}
  {\bibfnamefont {Q.}~\bibnamefont {Zhao}}, \bibinfo {author} {\bibfnamefont
  {J.~F.}\ \bibnamefont {Mitchell}}, \bibinfo {author} {\bibfnamefont {J.~W.}\
  \bibnamefont {Allen}}, \ and\ \bibinfo {author} {\bibfnamefont {B.~J.}\
  \bibnamefont {Kim}},\ }\href {\doibase 10.1126/science.1251151} {\bibfield
  {journal} {\bibinfo  {journal} {Science}\ }\textbf {\bibinfo {volume}
  {345}},\ \bibinfo {pages} {187} (\bibinfo {year} {2014})}\BibitemShut
  {NoStop}%
\bibitem [{\citenamefont {Cao}\ \emph {et~al.}(2014)\citenamefont {Cao},
  \citenamefont {Wang}, \citenamefont {Waugh}, \citenamefont {Reber},
  \citenamefont {Li}, \citenamefont {Zhou}, \citenamefont {Parham},
  \citenamefont {Plumb}, \citenamefont {Rotenberg}, \citenamefont {Bostwick},
  \citenamefont {Denlinger}, \citenamefont {Qi}, \citenamefont {Hermele},
  \citenamefont {Cao},\ and\ \citenamefont {Dessau}}]{Cao2014}%
  \BibitemOpen
  \bibfield  {author} {\bibinfo {author} {\bibfnamefont {Y.}~\bibnamefont
  {Cao}}, \bibinfo {author} {\bibfnamefont {Q.}~\bibnamefont {Wang}}, \bibinfo
  {author} {\bibfnamefont {H.}~\bibnamefont {Waugh}}, \bibinfo {author}
  {\bibfnamefont {T.~J.}\ \bibnamefont {Reber}}, \bibinfo {author}
  {\bibfnamefont {H.}~\bibnamefont {Li}}, \bibinfo {author} {\bibfnamefont
  {X.}~\bibnamefont {Zhou}}, \bibinfo {author} {\bibfnamefont {S.}~\bibnamefont
  {Parham}}, \bibinfo {author} {\bibfnamefont {N.~C.}\ \bibnamefont {Plumb}},
  \bibinfo {author} {\bibfnamefont {E.}~\bibnamefont {Rotenberg}}, \bibinfo
  {author} {\bibfnamefont {A.}~\bibnamefont {Bostwick}}, \bibinfo {author}
  {\bibfnamefont {D.}~\bibnamefont {Denlinger}}, \bibinfo {author}
  {\bibfnamefont {T.}~\bibnamefont {Qi}}, \bibinfo {author} {\bibfnamefont
  {A.}~\bibnamefont {Hermele}}, \bibinfo {author} {\bibfnamefont
  {G.}~\bibnamefont {Cao}}, \ and\ \bibinfo {author} {\bibfnamefont
  {D.}~\bibnamefont {Dessau}},\ }\href@noop {} {\bibfield  {journal} {\bibinfo
  {journal} {arXiv:1406.4978}\ } (\bibinfo {year} {2014})}\BibitemShut
  {NoStop}%
\bibitem [{\citenamefont {Norman}\ \emph {et~al.}(1998)\citenamefont {Norman},
  \citenamefont {Ding},\ and\ \citenamefont {Randeria}}]{Norman1998}%
  \BibitemOpen
  \bibfield  {author} {\bibinfo {author} {\bibfnamefont {M.}~\bibnamefont
  {Norman}}, \bibinfo {author} {\bibfnamefont {H.}~\bibnamefont {Ding}}, \ and\
  \bibinfo {author} {\bibfnamefont {M.}~\bibnamefont {Randeria}},\ }\href
  {http://www.nature.com/nature/journal/v392/n6672/abs/392157a0.html}
  {\bibfield  {journal} {\bibinfo  {journal} {Nature}\ }\textbf {\bibinfo
  {volume} {392}},\ \bibinfo {pages} {157} (\bibinfo {year}
  {1998})}\BibitemShut {NoStop}%
\bibitem [{\citenamefont {Shen}\ \emph {et~al.}(2005)\citenamefont {Shen},
  \citenamefont {Ronning}, \citenamefont {Lu}, \citenamefont {Baumberger},
  \citenamefont {Ingle}, \citenamefont {Lee}, \citenamefont {Meevasana},
  \citenamefont {Kohsaka}, \citenamefont {Azuma}, \citenamefont {Takano},
  \citenamefont {Takagi},\ and\ \citenamefont {Shen}}]{Shen2005}%
  \BibitemOpen
  \bibfield  {author} {\bibinfo {author} {\bibfnamefont {K.~M.}\ \bibnamefont
  {Shen}}, \bibinfo {author} {\bibfnamefont {F.}~\bibnamefont {Ronning}},
  \bibinfo {author} {\bibfnamefont {D.~H.}\ \bibnamefont {Lu}}, \bibinfo
  {author} {\bibfnamefont {F.}~\bibnamefont {Baumberger}}, \bibinfo {author}
  {\bibfnamefont {N.~J.~C.}\ \bibnamefont {Ingle}}, \bibinfo {author}
  {\bibfnamefont {W.~S.}\ \bibnamefont {Lee}}, \bibinfo {author} {\bibfnamefont
  {W.}~\bibnamefont {Meevasana}}, \bibinfo {author} {\bibfnamefont
  {Y.}~\bibnamefont {Kohsaka}}, \bibinfo {author} {\bibfnamefont
  {M.}~\bibnamefont {Azuma}}, \bibinfo {author} {\bibfnamefont
  {M.}~\bibnamefont {Takano}}, \bibinfo {author} {\bibfnamefont
  {H.}~\bibnamefont {Takagi}}, \ and\ \bibinfo {author} {\bibfnamefont {Z.~X.}\
  \bibnamefont {Shen}},\ }\href {<Go to ISI>://000226985100046} {\bibfield
  {journal} {\bibinfo  {journal} {Science}\ }\textbf {\bibinfo {volume}
  {307}},\ \bibinfo {pages} {901} (\bibinfo {year} {2005})}\BibitemShut
  {NoStop}%
\bibitem [{\citenamefont {Damascelli}\ \emph {et~al.}(2003)\citenamefont
  {Damascelli}, \citenamefont {Hussain},\ and\ \citenamefont
  {Shen}}]{Damascelli2003}%
  \BibitemOpen
  \bibfield  {author} {\bibinfo {author} {\bibfnamefont {A.}~\bibnamefont
  {Damascelli}}, \bibinfo {author} {\bibfnamefont {Z.}~\bibnamefont {Hussain}},
  \ and\ \bibinfo {author} {\bibfnamefont {Z.-X.}\ \bibnamefont {Shen}},\
  }\href {http://link.aps.org/abstract/RMP/v75/p473} {\bibfield  {journal}
  {\bibinfo  {journal} {Rev. Mod. Phys.}\ }\textbf {\bibinfo {volume} {75}},\
  \bibinfo {pages} {469} (\bibinfo {year} {2003})}\BibitemShut {NoStop}%
\bibitem [{\citenamefont {Dhital}\ \emph {et~al.}(2014)\citenamefont {Dhital},
  \citenamefont {Hogan}, \citenamefont {Zhou}, \citenamefont {Chen},
  \citenamefont {Ren}, \citenamefont {Pokharel}, \citenamefont {Okada},
  \citenamefont {Heine}, \citenamefont {Tian}, \citenamefont {Yamani},
  \citenamefont {Opeil}, \citenamefont {Helton}, \citenamefont {Lynn},
  \citenamefont {Wang}, \citenamefont {Madhavan},\ and\ \citenamefont
  {Wilson}}]{Dhital2014}%
  \BibitemOpen
  \bibfield  {author} {\bibinfo {author} {\bibfnamefont {C.}~\bibnamefont
  {Dhital}}, \bibinfo {author} {\bibfnamefont {T.}~\bibnamefont {Hogan}},
  \bibinfo {author} {\bibfnamefont {W.}~\bibnamefont {Zhou}}, \bibinfo {author}
  {\bibfnamefont {X.}~\bibnamefont {Chen}}, \bibinfo {author} {\bibfnamefont
  {Z.}~\bibnamefont {Ren}}, \bibinfo {author} {\bibfnamefont {M.}~\bibnamefont
  {Pokharel}}, \bibinfo {author} {\bibfnamefont {Y.}~\bibnamefont {Okada}},
  \bibinfo {author} {\bibfnamefont {M.}~\bibnamefont {Heine}}, \bibinfo
  {author} {\bibfnamefont {W.}~\bibnamefont {Tian}}, \bibinfo {author}
  {\bibfnamefont {Z.}~\bibnamefont {Yamani}}, \bibinfo {author} {\bibfnamefont
  {C.}~\bibnamefont {Opeil}}, \bibinfo {author} {\bibfnamefont {J.~S.}\
  \bibnamefont {Helton}}, \bibinfo {author} {\bibfnamefont {J.~W.}\
  \bibnamefont {Lynn}}, \bibinfo {author} {\bibfnamefont {Z.}~\bibnamefont
  {Wang}}, \bibinfo {author} {\bibfnamefont {V.}~\bibnamefont {Madhavan}}, \
  and\ \bibinfo {author} {\bibfnamefont {S.~D.}\ \bibnamefont {Wilson}},\
  }\href {\doibase 10.1038/ncomms4377} {\bibfield  {journal} {\bibinfo
  {journal} {Nat. Commun.}\ }\textbf {\bibinfo {volume} {5}},\ \bibinfo {pages}
  {3377} (\bibinfo {year} {2014})}\BibitemShut {NoStop}%
\bibitem [{\citenamefont {Boseggia}\ \emph {et~al.}(2012)\citenamefont
  {Boseggia}, \citenamefont {Springell}, \citenamefont {Walker}, \citenamefont
  {Boothroyd}, \citenamefont {Prabhakaran}, \citenamefont {Collins},\ and\
  \citenamefont {McMorrow}}]{Boseggia2012}%
  \BibitemOpen
  \bibfield  {author} {\bibinfo {author} {\bibfnamefont {S.}~\bibnamefont
  {Boseggia}}, \bibinfo {author} {\bibfnamefont {R.}~\bibnamefont {Springell}},
  \bibinfo {author} {\bibfnamefont {H.~C.}\ \bibnamefont {Walker}}, \bibinfo
  {author} {\bibfnamefont {A.~T.}\ \bibnamefont {Boothroyd}}, \bibinfo {author}
  {\bibfnamefont {D.}~\bibnamefont {Prabhakaran}}, \bibinfo {author}
  {\bibfnamefont {S.~P.}\ \bibnamefont {Collins}}, \ and\ \bibinfo {author}
  {\bibfnamefont {D.~F.}\ \bibnamefont {McMorrow}},\ }\href {\doibase
  10.1088/0953-8984/24/31/312202} {\bibfield  {journal} {\bibinfo  {journal}
  {J. Phys. Condens. Matter}\ }\textbf {\bibinfo {volume} {24}},\ \bibinfo
  {pages} {312202} (\bibinfo {year} {2012})}\BibitemShut {NoStop}%
\bibitem [{\citenamefont {Kim}\ \emph {et~al.}(2012{\natexlab{b}})\citenamefont
  {Kim}, \citenamefont {Choi}, \citenamefont {Kim}, \citenamefont {Mitchell},
  \citenamefont {Jackeli}, \citenamefont {Daghofer}, \citenamefont {van~den
  Brink}, \citenamefont {Khaliullin},\ and\ \citenamefont {Kim}}]{Kim2012}%
  \BibitemOpen
  \bibfield  {author} {\bibinfo {author} {\bibfnamefont {J.~W.}\ \bibnamefont
  {Kim}}, \bibinfo {author} {\bibfnamefont {Y.}~\bibnamefont {Choi}}, \bibinfo
  {author} {\bibfnamefont {J.}~\bibnamefont {Kim}}, \bibinfo {author}
  {\bibfnamefont {J.~F.}\ \bibnamefont {Mitchell}}, \bibinfo {author}
  {\bibfnamefont {G.}~\bibnamefont {Jackeli}}, \bibinfo {author} {\bibfnamefont
  {M.}~\bibnamefont {Daghofer}}, \bibinfo {author} {\bibfnamefont
  {J.}~\bibnamefont {van~den Brink}}, \bibinfo {author} {\bibfnamefont
  {G.}~\bibnamefont {Khaliullin}}, \ and\ \bibinfo {author} {\bibfnamefont
  {B.~J.}\ \bibnamefont {Kim}},\ }\href {\doibase
  10.1103/PhysRevLett.109.037204} {\bibfield  {journal} {\bibinfo  {journal}
  {Phys. Rev. Lett.}\ }\textbf {\bibinfo {volume} {109}},\ \bibinfo {pages}
  {037204} (\bibinfo {year} {2012}{\natexlab{b}})}\BibitemShut {NoStop}%
\bibitem [{\citenamefont {Kim}\ \emph {et~al.}(2012{\natexlab{c}})\citenamefont
  {Kim}, \citenamefont {Said}, \citenamefont {Casa}, \citenamefont {Upton},
  \citenamefont {Gog}, \citenamefont {Daghofer}, \citenamefont {Jackeli},
  \citenamefont {van~den Brink}, \citenamefont {Khaliullin},\ and\
  \citenamefont {Kim}}]{Kim2012b}%
  \BibitemOpen
  \bibfield  {author} {\bibinfo {author} {\bibfnamefont {J.}~\bibnamefont
  {Kim}}, \bibinfo {author} {\bibfnamefont {A.~H.}\ \bibnamefont {Said}},
  \bibinfo {author} {\bibfnamefont {D.}~\bibnamefont {Casa}}, \bibinfo {author}
  {\bibfnamefont {M.~H.}\ \bibnamefont {Upton}}, \bibinfo {author}
  {\bibfnamefont {T.}~\bibnamefont {Gog}}, \bibinfo {author} {\bibfnamefont
  {M.}~\bibnamefont {Daghofer}}, \bibinfo {author} {\bibfnamefont
  {G.}~\bibnamefont {Jackeli}}, \bibinfo {author} {\bibfnamefont
  {J.}~\bibnamefont {van~den Brink}}, \bibinfo {author} {\bibfnamefont
  {G.}~\bibnamefont {Khaliullin}}, \ and\ \bibinfo {author} {\bibfnamefont
  {B.~J.}\ \bibnamefont {Kim}},\ }\href {\doibase
  10.1103/PhysRevLett.109.157402} {\bibfield  {journal} {\bibinfo  {journal}
  {Phys. Rev. Lett.}\ }\textbf {\bibinfo {volume} {109}},\ \bibinfo {pages}
  {157402} (\bibinfo {year} {2012}{\natexlab{c}})}\BibitemShut {NoStop}%
\bibitem [{\citenamefont {Okada}\ \emph {et~al.}(2013)\citenamefont {Okada},
  \citenamefont {Walkup}, \citenamefont {Lin}, \citenamefont {Dhital},
  \citenamefont {Chang}, \citenamefont {Khadka}, \citenamefont {Zhou},
  \citenamefont {Jeng}, \citenamefont {Paranjape}, \citenamefont {Bansil},
  \citenamefont {Wang}, \citenamefont {Wilson},\ and\ \citenamefont
  {Madhavan}}]{Okada2013}%
  \BibitemOpen
  \bibfield  {author} {\bibinfo {author} {\bibfnamefont {Y.}~\bibnamefont
  {Okada}}, \bibinfo {author} {\bibfnamefont {D.}~\bibnamefont {Walkup}},
  \bibinfo {author} {\bibfnamefont {H.}~\bibnamefont {Lin}}, \bibinfo {author}
  {\bibfnamefont {C.}~\bibnamefont {Dhital}}, \bibinfo {author} {\bibfnamefont
  {T.-R.}\ \bibnamefont {Chang}}, \bibinfo {author} {\bibfnamefont
  {S.}~\bibnamefont {Khadka}}, \bibinfo {author} {\bibfnamefont
  {W.}~\bibnamefont {Zhou}}, \bibinfo {author} {\bibfnamefont {H.-T.}\
  \bibnamefont {Jeng}}, \bibinfo {author} {\bibfnamefont {M.}~\bibnamefont
  {Paranjape}}, \bibinfo {author} {\bibfnamefont {A.}~\bibnamefont {Bansil}},
  \bibinfo {author} {\bibfnamefont {Z.}~\bibnamefont {Wang}}, \bibinfo {author}
  {\bibfnamefont {S.~D.}\ \bibnamefont {Wilson}}, \ and\ \bibinfo {author}
  {\bibfnamefont {V.}~\bibnamefont {Madhavan}},\ }\href {\doibase
  10.1038/nmat3653} {\bibfield  {journal} {\bibinfo  {journal} {Nat. Mater.}\
  }\textbf {\bibinfo {volume} {12}},\ \bibinfo {pages} {707} (\bibinfo {year}
  {2013})}\BibitemShut {NoStop}%
\bibitem [{\citenamefont {Moon}\ \emph {et~al.}(2008)\citenamefont {Moon},
  \citenamefont {Jin}, \citenamefont {Kim}, \citenamefont {Choi}, \citenamefont
  {Lee}, \citenamefont {Yu}, \citenamefont {Cao}, \citenamefont {Sumi},
  \citenamefont {Funakubo}, \citenamefont {Bernhard},\ and\ \citenamefont
  {Noh}}]{Moon2008}%
  \BibitemOpen
  \bibfield  {author} {\bibinfo {author} {\bibfnamefont {S.~J.}\ \bibnamefont
  {Moon}}, \bibinfo {author} {\bibfnamefont {H.}~\bibnamefont {Jin}}, \bibinfo
  {author} {\bibfnamefont {K.~W.}\ \bibnamefont {Kim}}, \bibinfo {author}
  {\bibfnamefont {W.~S.}\ \bibnamefont {Choi}}, \bibinfo {author}
  {\bibfnamefont {Y.~S.}\ \bibnamefont {Lee}}, \bibinfo {author} {\bibfnamefont
  {J.}~\bibnamefont {Yu}}, \bibinfo {author} {\bibfnamefont {G.}~\bibnamefont
  {Cao}}, \bibinfo {author} {\bibfnamefont {A.}~\bibnamefont {Sumi}}, \bibinfo
  {author} {\bibfnamefont {H.}~\bibnamefont {Funakubo}}, \bibinfo {author}
  {\bibfnamefont {C.}~\bibnamefont {Bernhard}}, \ and\ \bibinfo {author}
  {\bibfnamefont {T.~W.}\ \bibnamefont {Noh}},\ }\href
  {http://link.aps.org/abstract/PRL/v101/e226402} {\bibfield  {journal}
  {\bibinfo  {journal} {Phys. Rev. Lett.}\ }\textbf {\bibinfo {volume} {101}},\
  \bibinfo {pages} {226402} (\bibinfo {year} {2008})}\BibitemShut {NoStop}%
\bibitem [{\citenamefont {Fujiyama}\ \emph
  {et~al.}(2012{\natexlab{b}})\citenamefont {Fujiyama}, \citenamefont {Ohashi},
  \citenamefont {Ohsumi}, \citenamefont {Sugimoto}, \citenamefont {Takayama},
  \citenamefont {Komesu}, \citenamefont {Takata}, \citenamefont {Arima},\ and\
  \citenamefont {Takagi}}]{Fujiyama2012}%
  \BibitemOpen
  \bibfield  {author} {\bibinfo {author} {\bibfnamefont {S.}~\bibnamefont
  {Fujiyama}}, \bibinfo {author} {\bibfnamefont {K.}~\bibnamefont {Ohashi}},
  \bibinfo {author} {\bibfnamefont {H.}~\bibnamefont {Ohsumi}}, \bibinfo
  {author} {\bibfnamefont {K.}~\bibnamefont {Sugimoto}}, \bibinfo {author}
  {\bibfnamefont {T.}~\bibnamefont {Takayama}}, \bibinfo {author}
  {\bibfnamefont {T.}~\bibnamefont {Komesu}}, \bibinfo {author} {\bibfnamefont
  {M.}~\bibnamefont {Takata}}, \bibinfo {author} {\bibfnamefont
  {T.}~\bibnamefont {Arima}}, \ and\ \bibinfo {author} {\bibfnamefont
  {H.}~\bibnamefont {Takagi}},\ }\href {\doibase 10.1103/PhysRevB.86.174414}
  {\bibfield  {journal} {\bibinfo  {journal} {Phys. Rev. B}\ }\textbf {\bibinfo
  {volume} {86}},\ \bibinfo {pages} {174414} (\bibinfo {year}
  {2012}{\natexlab{b}})}\BibitemShut {NoStop}%
\bibitem [{\citenamefont {Park}\ \emph {et~al.}(2014)\citenamefont {Park},
  \citenamefont {Sohn}, \citenamefont {Jeong}, \citenamefont {Cao},
  \citenamefont {Kim}, \citenamefont {Moon}, \citenamefont {Jin}, \citenamefont
  {Cho},\ and\ \citenamefont {Noh}}]{Park2014}%
  \BibitemOpen
  \bibfield  {author} {\bibinfo {author} {\bibfnamefont {H.~J.}\ \bibnamefont
  {Park}}, \bibinfo {author} {\bibfnamefont {C.~H.}\ \bibnamefont {Sohn}},
  \bibinfo {author} {\bibfnamefont {D.~W.}\ \bibnamefont {Jeong}}, \bibinfo
  {author} {\bibfnamefont {G.}~\bibnamefont {Cao}}, \bibinfo {author}
  {\bibfnamefont {K.~W.}\ \bibnamefont {Kim}}, \bibinfo {author} {\bibfnamefont
  {S.~J.}\ \bibnamefont {Moon}}, \bibinfo {author} {\bibfnamefont
  {H.}~\bibnamefont {Jin}}, \bibinfo {author} {\bibfnamefont {D.-Y.}\
  \bibnamefont {Cho}}, \ and\ \bibinfo {author} {\bibfnamefont {T.~W.}\
  \bibnamefont {Noh}},\ }\href {\doibase 10.1103/PhysRevB.89.155115} {\bibfield
   {journal} {\bibinfo  {journal} {Phys. Rev. B}\ }\textbf {\bibinfo {volume}
  {89}},\ \bibinfo {pages} {155115} (\bibinfo {year} {2014})}\BibitemShut
  {NoStop}%
\bibitem [{\citenamefont {Wojek}\ \emph {et~al.}(2012)\citenamefont {Wojek},
  \citenamefont {Berntsen}, \citenamefont {Boseggia}, \citenamefont
  {Boothroyd}, \citenamefont {Prabhakaran}, \citenamefont {McMorrow},
  \citenamefont {R{\o}nnow}, \citenamefont {Chang},\ and\ \citenamefont
  {Tjernberg}}]{Wojek2012}%
  \BibitemOpen
  \bibfield  {author} {\bibinfo {author} {\bibfnamefont {B.~M.}\ \bibnamefont
  {Wojek}}, \bibinfo {author} {\bibfnamefont {M.~H.}\ \bibnamefont {Berntsen}},
  \bibinfo {author} {\bibfnamefont {S.}~\bibnamefont {Boseggia}}, \bibinfo
  {author} {\bibfnamefont {A.~T.}\ \bibnamefont {Boothroyd}}, \bibinfo {author}
  {\bibfnamefont {D.}~\bibnamefont {Prabhakaran}}, \bibinfo {author}
  {\bibfnamefont {D.~F.}\ \bibnamefont {McMorrow}}, \bibinfo {author}
  {\bibfnamefont {H.~M.}\ \bibnamefont {R{\o}nnow}}, \bibinfo {author}
  {\bibfnamefont {J.}~\bibnamefont {Chang}}, \ and\ \bibinfo {author}
  {\bibfnamefont {O.}~\bibnamefont {Tjernberg}},\ }\href {\doibase
  10.1088/0953-8984/24/41/415602} {\bibfield  {journal} {\bibinfo  {journal}
  {J. Phys. Condens. Matter}\ }\textbf {\bibinfo {volume} {24}},\ \bibinfo
  {pages} {415602} (\bibinfo {year} {2012})}\BibitemShut {NoStop}%
\bibitem [{\citenamefont {King}\ \emph {et~al.}(2013)\citenamefont {King},
  \citenamefont {Takayama}, \citenamefont {Tamai}, \citenamefont {Rozbicki},
  \citenamefont {Walker}, \citenamefont {Shi}, \citenamefont {Patthey},
  \citenamefont {Moore}, \citenamefont {Lu}, \citenamefont {Shen},
  \citenamefont {Takagi},\ and\ \citenamefont {Baumberger}}]{King2013}%
  \BibitemOpen
  \bibfield  {author} {\bibinfo {author} {\bibfnamefont {P.~D.~C.}\
  \bibnamefont {King}}, \bibinfo {author} {\bibfnamefont {T.}~\bibnamefont
  {Takayama}}, \bibinfo {author} {\bibfnamefont {A.}~\bibnamefont {Tamai}},
  \bibinfo {author} {\bibfnamefont {E.}~\bibnamefont {Rozbicki}}, \bibinfo
  {author} {\bibfnamefont {S.~M.}\ \bibnamefont {Walker}}, \bibinfo {author}
  {\bibfnamefont {M.}~\bibnamefont {Shi}}, \bibinfo {author} {\bibfnamefont
  {L.}~\bibnamefont {Patthey}}, \bibinfo {author} {\bibfnamefont {R.~G.}\
  \bibnamefont {Moore}}, \bibinfo {author} {\bibfnamefont {D.}~\bibnamefont
  {Lu}}, \bibinfo {author} {\bibfnamefont {K.~M.}\ \bibnamefont {Shen}},
  \bibinfo {author} {\bibfnamefont {H.}~\bibnamefont {Takagi}}, \ and\ \bibinfo
  {author} {\bibfnamefont {F.}~\bibnamefont {Baumberger}},\ }\href {\doibase
  10.1103/PhysRevB.87.241106} {\bibfield  {journal} {\bibinfo  {journal} {Phys.
  Rev. B}\ }\textbf {\bibinfo {volume} {87}},\ \bibinfo {pages} {241106}
  (\bibinfo {year} {2013})}\BibitemShut {NoStop}%
\bibitem [{\citenamefont {Wang}\ \emph {et~al.}(2013)\citenamefont {Wang},
  \citenamefont {Cao}, \citenamefont {Waugh}, \citenamefont {Park},
  \citenamefont {Qi}, \citenamefont {Korneta}, \citenamefont {Cao},\ and\
  \citenamefont {Dessau}}]{Wang2013}%
  \BibitemOpen
  \bibfield  {author} {\bibinfo {author} {\bibfnamefont {Q.}~\bibnamefont
  {Wang}}, \bibinfo {author} {\bibfnamefont {Y.}~\bibnamefont {Cao}}, \bibinfo
  {author} {\bibfnamefont {J.~A.}\ \bibnamefont {Waugh}}, \bibinfo {author}
  {\bibfnamefont {S.~R.}\ \bibnamefont {Park}}, \bibinfo {author}
  {\bibfnamefont {T.~F.}\ \bibnamefont {Qi}}, \bibinfo {author} {\bibfnamefont
  {O.~B.}\ \bibnamefont {Korneta}}, \bibinfo {author} {\bibfnamefont
  {G.}~\bibnamefont {Cao}}, \ and\ \bibinfo {author} {\bibfnamefont {D.~S.}\
  \bibnamefont {Dessau}},\ }\href {\doibase 10.1103/PhysRevB.87.245109}
  {\bibfield  {journal} {\bibinfo  {journal} {Phys. Rev. B}\ }\textbf {\bibinfo
  {volume} {87}},\ \bibinfo {pages} {245109} (\bibinfo {year}
  {2013})}\BibitemShut {NoStop}%
\bibitem [{\citenamefont {Moreschini}\ \emph {et~al.}(2014)\citenamefont
  {Moreschini}, \citenamefont {Moser}, \citenamefont {Ebrahimi}, \citenamefont
  {{Dalla Piazza}}, \citenamefont {Kim}, \citenamefont {Boseggia},
  \citenamefont {McMorrow}, \citenamefont {R{\o}nnow}, \citenamefont {Chang},
  \citenamefont {Prabhakaran}, \citenamefont {Boothroyd}, \citenamefont
  {Rotenberg}, \citenamefont {Bostwick},\ and\ \citenamefont
  {Grioni}}]{Moreschini2014}%
  \BibitemOpen
  \bibfield  {author} {\bibinfo {author} {\bibfnamefont {L.}~\bibnamefont
  {Moreschini}}, \bibinfo {author} {\bibfnamefont {S.}~\bibnamefont {Moser}},
  \bibinfo {author} {\bibfnamefont {A.}~\bibnamefont {Ebrahimi}}, \bibinfo
  {author} {\bibfnamefont {B.}~\bibnamefont {{Dalla Piazza}}}, \bibinfo
  {author} {\bibfnamefont {K.~S.}\ \bibnamefont {Kim}}, \bibinfo {author}
  {\bibfnamefont {S.}~\bibnamefont {Boseggia}}, \bibinfo {author}
  {\bibfnamefont {D.~F.}\ \bibnamefont {McMorrow}}, \bibinfo {author}
  {\bibfnamefont {H.~M.}\ \bibnamefont {R{\o}nnow}}, \bibinfo {author}
  {\bibfnamefont {J.}~\bibnamefont {Chang}}, \bibinfo {author} {\bibfnamefont
  {D.}~\bibnamefont {Prabhakaran}}, \bibinfo {author} {\bibfnamefont {A.~T.}\
  \bibnamefont {Boothroyd}}, \bibinfo {author} {\bibfnamefont {E.}~\bibnamefont
  {Rotenberg}}, \bibinfo {author} {\bibfnamefont {A.}~\bibnamefont {Bostwick}},
  \ and\ \bibinfo {author} {\bibfnamefont {M.}~\bibnamefont {Grioni}},\ }\href
  {\doibase 10.1103/PhysRevB.89.201114} {\bibfield  {journal} {\bibinfo
  {journal} {Phys. Rev. B}\ }\textbf {\bibinfo {volume} {89}},\ \bibinfo
  {pages} {201114} (\bibinfo {year} {2014})}\BibitemShut {NoStop}%
\bibitem [{\citenamefont {Li}\ \emph {et~al.}(2013)\citenamefont {Li},
  \citenamefont {Kong}, \citenamefont {Qi}, \citenamefont {Jin}, \citenamefont
  {Yuan}, \citenamefont {DeLong}, \citenamefont {Schlottmann},\ and\
  \citenamefont {Cao}}]{Li2013}%
  \BibitemOpen
  \bibfield  {author} {\bibinfo {author} {\bibfnamefont {L.}~\bibnamefont
  {Li}}, \bibinfo {author} {\bibfnamefont {P.~P.}\ \bibnamefont {Kong}},
  \bibinfo {author} {\bibfnamefont {T.~F.}\ \bibnamefont {Qi}}, \bibinfo
  {author} {\bibfnamefont {C.~Q.}\ \bibnamefont {Jin}}, \bibinfo {author}
  {\bibfnamefont {S.~J.}\ \bibnamefont {Yuan}}, \bibinfo {author}
  {\bibfnamefont {L.~E.}\ \bibnamefont {DeLong}}, \bibinfo {author}
  {\bibfnamefont {P.}~\bibnamefont {Schlottmann}}, \ and\ \bibinfo {author}
  {\bibfnamefont {G.}~\bibnamefont {Cao}},\ }\href {\doibase
  10.1103/PhysRevB.87.235127} {\bibfield  {journal} {\bibinfo  {journal} {Phys.
  Rev. B}\ }\textbf {\bibinfo {volume} {87}},\ \bibinfo {pages} {235127}
  (\bibinfo {year} {2013})}\BibitemShut {NoStop}%
\bibitem [{\citenamefont {Hunter}\ and\ \citenamefont
  {Perry}(2014)}]{Hunter2014}%
  \BibitemOpen
  \bibfield  {author} {\bibinfo {author} {\bibfnamefont {E.~C.}\ \bibnamefont
  {Hunter}}\ and\ \bibinfo {author} {\bibfnamefont {R.~S.}\ \bibnamefont
  {Perry}},\ }\href@noop {} {\bibfield  {journal} {\bibinfo  {journal} {in
  preparation}\ } (\bibinfo {year} {2014})}\BibitemShut {NoStop}%
\bibitem [{\citenamefont {Blaha}\ \emph {et~al.}(2001)\citenamefont {Blaha},
  \citenamefont {Schwarz}, \citenamefont {Madsen}, \citenamefont {Kvasnicka},\
  and\ \citenamefont {Luitz}}]{Blaha2001}%
  \BibitemOpen
  \bibfield  {author} {\bibinfo {author} {\bibfnamefont {P.}~\bibnamefont
  {Blaha}}, \bibinfo {author} {\bibfnamefont {K.}~\bibnamefont {Schwarz}},
  \bibinfo {author} {\bibfnamefont {G.}~\bibnamefont {Madsen}}, \bibinfo
  {author} {\bibfnamefont {D.}~\bibnamefont {Kvasnicka}}, \ and\ \bibinfo
  {author} {\bibfnamefont {J.}~\bibnamefont {Luitz}},\ }\href@noop {} {\emph
  {\bibinfo {title} {WIEN2k, An Augmented Plane Wave + Local Orbitals Program
  for Calculating Crystal Properties}}}\ (\bibinfo  {publisher} {K. Schwarz,
  Tech. Univ. Wien, Austria},\ \bibinfo {year} {2001})\BibitemShut {NoStop}%
\bibitem [{\citenamefont {Ino}\ \emph {et~al.}(2002)\citenamefont {Ino},
  \citenamefont {Kim}, \citenamefont {Nakamura}, \citenamefont {Yoshida},
  \citenamefont {Mizokawa}, \citenamefont {Fujimori}, \citenamefont {Shen},
  \citenamefont {Kakeshita}, \citenamefont {Eisaki},\ and\ \citenamefont
  {Uchida}}]{Ino2002}%
  \BibitemOpen
  \bibfield  {author} {\bibinfo {author} {\bibfnamefont {A.}~\bibnamefont
  {Ino}}, \bibinfo {author} {\bibfnamefont {C.}~\bibnamefont {Kim}}, \bibinfo
  {author} {\bibfnamefont {M.}~\bibnamefont {Nakamura}}, \bibinfo {author}
  {\bibfnamefont {T.}~\bibnamefont {Yoshida}}, \bibinfo {author} {\bibfnamefont
  {T.}~\bibnamefont {Mizokawa}}, \bibinfo {author} {\bibfnamefont
  {A.}~\bibnamefont {Fujimori}}, \bibinfo {author} {\bibfnamefont {Z.-X.}\
  \bibnamefont {Shen}}, \bibinfo {author} {\bibfnamefont {T.}~\bibnamefont
  {Kakeshita}}, \bibinfo {author} {\bibfnamefont {H.}~\bibnamefont {Eisaki}}, \
  and\ \bibinfo {author} {\bibfnamefont {S.}~\bibnamefont {Uchida}},\ }\href
  {\doibase 10.1103/PhysRevB.65.094504} {\bibfield  {journal} {\bibinfo
  {journal} {Phys. Rev. B}\ }\textbf {\bibinfo {volume} {65}},\ \bibinfo
  {pages} {094504} (\bibinfo {year} {2002})}\BibitemShut {NoStop}%
\bibitem [{\citenamefont {Shen}\ \emph {et~al.}(2004)\citenamefont {Shen},
  \citenamefont {Ronning}, \citenamefont {Lu}, \citenamefont {Lee},
  \citenamefont {Ingle}, \citenamefont {Meevasana}, \citenamefont {Baumberger},
  \citenamefont {Damascelli}, \citenamefont {Armitage}, \citenamefont {Miller},
  \citenamefont {Kohsaka}, \citenamefont {Azuma}, \citenamefont {Takano},
  \citenamefont {Takagi},\ and\ \citenamefont {Shen}}]{Shen2004}%
  \BibitemOpen
  \bibfield  {author} {\bibinfo {author} {\bibfnamefont {K.~M.}\ \bibnamefont
  {Shen}}, \bibinfo {author} {\bibfnamefont {F.}~\bibnamefont {Ronning}},
  \bibinfo {author} {\bibfnamefont {D.~H.}\ \bibnamefont {Lu}}, \bibinfo
  {author} {\bibfnamefont {W.~S.}\ \bibnamefont {Lee}}, \bibinfo {author}
  {\bibfnamefont {N.~J.~C.}\ \bibnamefont {Ingle}}, \bibinfo {author}
  {\bibfnamefont {W.}~\bibnamefont {Meevasana}}, \bibinfo {author}
  {\bibfnamefont {F.}~\bibnamefont {Baumberger}}, \bibinfo {author}
  {\bibfnamefont {A.}~\bibnamefont {Damascelli}}, \bibinfo {author}
  {\bibfnamefont {N.~P.}\ \bibnamefont {Armitage}}, \bibinfo {author}
  {\bibfnamefont {L.~L.}\ \bibnamefont {Miller}}, \bibinfo {author}
  {\bibfnamefont {Y.}~\bibnamefont {Kohsaka}}, \bibinfo {author} {\bibfnamefont
  {M.}~\bibnamefont {Azuma}}, \bibinfo {author} {\bibfnamefont
  {M.}~\bibnamefont {Takano}}, \bibinfo {author} {\bibfnamefont
  {H.}~\bibnamefont {Takagi}}, \ and\ \bibinfo {author} {\bibfnamefont {Z.~X.}\
  \bibnamefont {Shen}},\ }\href@noop {} {\bibfield  {journal} {\bibinfo
  {journal} {Phys. Rev. Lett.}\ }\textbf {\bibinfo {volume} {93}},\ \bibinfo
  {pages} {267002} (\bibinfo {year} {2004})}\BibitemShut {NoStop}%
\bibitem [{\citenamefont {Yoshida}\ \emph {et~al.}(2006)\citenamefont
  {Yoshida}, \citenamefont {Zhou}, \citenamefont {Tanaka}, \citenamefont
  {Yang}, \citenamefont {Hussain}, \citenamefont {Shen}, \citenamefont
  {Fujimori}, \citenamefont {Sahrakorpi}, \citenamefont {Lindroos},
  \citenamefont {Markiewicz}, \citenamefont {Bansil}, \citenamefont {Komiya},
  \citenamefont {Ando}, \citenamefont {Eisaki}, \citenamefont {Kakeshita},\
  and\ \citenamefont {Uchida}}]{Yoshida2006}%
  \BibitemOpen
  \bibfield  {author} {\bibinfo {author} {\bibfnamefont {T.}~\bibnamefont
  {Yoshida}}, \bibinfo {author} {\bibfnamefont {X.~J.}\ \bibnamefont {Zhou}},
  \bibinfo {author} {\bibfnamefont {K.}~\bibnamefont {Tanaka}}, \bibinfo
  {author} {\bibfnamefont {W.~L.}\ \bibnamefont {Yang}}, \bibinfo {author}
  {\bibfnamefont {Z.}~\bibnamefont {Hussain}}, \bibinfo {author} {\bibfnamefont
  {Z.-X.}\ \bibnamefont {Shen}}, \bibinfo {author} {\bibfnamefont
  {A.}~\bibnamefont {Fujimori}}, \bibinfo {author} {\bibfnamefont
  {S.}~\bibnamefont {Sahrakorpi}}, \bibinfo {author} {\bibfnamefont
  {M.}~\bibnamefont {Lindroos}}, \bibinfo {author} {\bibfnamefont {R.~S.}\
  \bibnamefont {Markiewicz}}, \bibinfo {author} {\bibfnamefont
  {A.}~\bibnamefont {Bansil}}, \bibinfo {author} {\bibfnamefont
  {S.}~\bibnamefont {Komiya}}, \bibinfo {author} {\bibfnamefont
  {Y.}~\bibnamefont {Ando}}, \bibinfo {author} {\bibfnamefont {H.}~\bibnamefont
  {Eisaki}}, \bibinfo {author} {\bibfnamefont {T.}~\bibnamefont {Kakeshita}}, \
  and\ \bibinfo {author} {\bibfnamefont {S.}~\bibnamefont {Uchida}},\ }\href
  {\doibase 10.1103/PhysRevB.74.224510} {\bibfield  {journal} {\bibinfo
  {journal} {Phys. Rev. B}\ }\textbf {\bibinfo {volume} {74}},\ \bibinfo
  {pages} {224510} (\bibinfo {year} {2006})}\BibitemShut {NoStop}%
\bibitem [{\citenamefont {Tamai}\ \emph {et~al.}(2008)\citenamefont {Tamai},
  \citenamefont {Allan}, \citenamefont {Mercure}, \citenamefont {Meevasana},
  \citenamefont {Dunkel}, \citenamefont {Lu}, \citenamefont {Perry},
  \citenamefont {Mackenzie}, \citenamefont {Singh}, \citenamefont {Shen},\ and\
  \citenamefont {Baumberger}}]{Tamai2008}%
  \BibitemOpen
  \bibfield  {author} {\bibinfo {author} {\bibfnamefont {A.}~\bibnamefont
  {Tamai}}, \bibinfo {author} {\bibfnamefont {M.~P.}\ \bibnamefont {Allan}},
  \bibinfo {author} {\bibfnamefont {J.~F.}\ \bibnamefont {Mercure}}, \bibinfo
  {author} {\bibfnamefont {W.}~\bibnamefont {Meevasana}}, \bibinfo {author}
  {\bibfnamefont {R.}~\bibnamefont {Dunkel}}, \bibinfo {author} {\bibfnamefont
  {D.~H.}\ \bibnamefont {Lu}}, \bibinfo {author} {\bibfnamefont {R.~S.}\
  \bibnamefont {Perry}}, \bibinfo {author} {\bibfnamefont {A.~P.}\ \bibnamefont
  {Mackenzie}}, \bibinfo {author} {\bibfnamefont {D.~J.}\ \bibnamefont
  {Singh}}, \bibinfo {author} {\bibfnamefont {Z.~X.}\ \bibnamefont {Shen}}, \
  and\ \bibinfo {author} {\bibfnamefont {F.}~\bibnamefont {Baumberger}},\
  }\href {http://link.aps.org/abstract/PRL/v101/e026407} {\bibfield  {journal}
  {\bibinfo  {journal} {Phys. Rev. Lett.}\ }\textbf {\bibinfo {volume} {101}},\
  \bibinfo {pages} {026407} (\bibinfo {year} {2008})}\BibitemShut {NoStop}%
\bibitem [{\citenamefont {Baumberger}\ \emph {et~al.}(2006)\citenamefont
  {Baumberger}, \citenamefont {Ingle}, \citenamefont {Meevasana}, \citenamefont
  {Shen}, \citenamefont {Lu}, \citenamefont {Perry}, \citenamefont {Mackenzie},
  \citenamefont {Hussain}, \citenamefont {Singh},\ and\ \citenamefont
  {Shen}}]{Baumberger2006}%
  \BibitemOpen
  \bibfield  {author} {\bibinfo {author} {\bibfnamefont {F.}~\bibnamefont
  {Baumberger}}, \bibinfo {author} {\bibfnamefont {N.~J.~C.}\ \bibnamefont
  {Ingle}}, \bibinfo {author} {\bibfnamefont {W.}~\bibnamefont {Meevasana}},
  \bibinfo {author} {\bibfnamefont {K.~M.}\ \bibnamefont {Shen}}, \bibinfo
  {author} {\bibfnamefont {D.~H.}\ \bibnamefont {Lu}}, \bibinfo {author}
  {\bibfnamefont {R.~S.}\ \bibnamefont {Perry}}, \bibinfo {author}
  {\bibfnamefont {A.~P.}\ \bibnamefont {Mackenzie}}, \bibinfo {author}
  {\bibfnamefont {Z.}~\bibnamefont {Hussain}}, \bibinfo {author} {\bibfnamefont
  {D.~J.}\ \bibnamefont {Singh}}, \ and\ \bibinfo {author} {\bibfnamefont
  {Z.~X.}\ \bibnamefont {Shen}},\ }\href@noop {} {\bibfield  {journal}
  {\bibinfo  {journal} {Phys. Rev. Lett.}\ }\textbf {\bibinfo {volume} {96}},\
  \bibinfo {pages} {246402} (\bibinfo {year} {2006})}\BibitemShut {NoStop}%
\bibitem [{\citenamefont {Georges}\ \emph {et~al.}(2013)\citenamefont
  {Georges}, \citenamefont {Medici},\ and\ \citenamefont
  {Mravlje}}]{Georges2013}%
  \BibitemOpen
  \bibfield  {author} {\bibinfo {author} {\bibfnamefont {A.}~\bibnamefont
  {Georges}}, \bibinfo {author} {\bibfnamefont {L.~D.}\ \bibnamefont {Medici}},
  \ and\ \bibinfo {author} {\bibfnamefont {J.}~\bibnamefont {Mravlje}},\ }\href
  {\doibase 10.1146/annurev-conmatphys-020911-125045} {\bibfield  {journal}
  {\bibinfo  {journal} {Annu. Rev. Condens. Matter Phys.}\ }\textbf {\bibinfo
  {volume} {4}},\ \bibinfo {pages} {137} (\bibinfo {year} {2013})}\BibitemShut
  {NoStop}%
\end{thebibliography}%

\end{document}